\documentclass[10pt]{iopart}
\usepackage{amssymb,amsthm}
\usepackage{a4} 
\usepackage[mathscr]{eucal}
\theoremstyle{plain}

 \long\def\symbolfootnote[#1]#2{\begingroup\def\thefootnote{\fnsymbol{footnote}}\footnote[#1]{#2}\endgroup} 
\renewcommand{\thefootnote}{\arabic{footnote}}
\usepackage{perpage} 
\MakePerPage{footnote}

\begin{document}

\title[Generalized relativistic hydrodynamics]{Generalized relativistic hydrodynamics
\\
with a convex extension\footnotetext{Published in Classical Quantum Gravity (2014).}}

\author{Robert Beig}

\address{Gravitational Physics, Faculty of Physics, University of Vienna, Boltzmanngasse 5, A-1090 Vienna, Austria}
\ead{robert.beig@univie.ac.at}

 \author{Philippe G. LeFloch} 

%\address{Philippe G. L{\scriptsize e}Floch, 

\address{Laboratoire Jacques-Louis Lions and Centre National de la Recherche Scientifique, Universit\'e Pierre et Marie Curie (Paris VI), 4 Place Jussieu, 75252 Paris, France}
\ead{contact@philippelefloch.org}

\begin{abstract}  We propose a new model which describes relativistic hydrodynamics and generalizes the standard Euler system of isentropic perfect fluids. Remarkably, our system admits a convex extension which allows us to transform it to a symmetric hyperbolic form. This result sheds new light even on the relativistic Euler system.
\end{abstract}


%================================================================

\date{}

\section{Introduction}

Relativistic hydrodynamics is classically based on the Euler system of perfect compressible fluids, which enjoys many properties such as the existence of a symmetrization and, consequently, an initial value formulation; see \cite{mu1,mu2,frau,walton,lfl,lr}. Our purpose in the present paper is to introduce a generalization of this model, which enjoys essentially the same algebraic and analytical structures, while possibly encompassing a larger class of fluid behaviors.\\
Our model, in short, is as follows: we require a tensor $T_\mu{}^\nu$ on $\mathbb{R}^4$ (viewed as spacetime) of the form
\begin{equation}\label{new}
T_\mu{}^\nu = \rho\, \delta_\mu{}^\nu + 2 \frac{\partial \rho}{\partial J^\alpha} \,\delta_\mu{}^{[\alpha} J^{\nu]}.
\end{equation}
Here the `energy density' $\rho$ is given as a function of $J^\mu$, which in turn plays the role of the `current density' on spacetime. The field equations are then given by the system of conservation laws
\begin{equation}\label{model}
\partial_\nu T_\mu{}^\nu = 0.
\end{equation}
The equations (\ref{new})-(\ref{model}) reduce to the relativistic Euler system when $\rho$ depends on $J^\mu$ only via the particle number density $n = (- \gamma_{\mu\nu} J^\mu J^\nu)^\frac{1}{2}$, where $\gamma_{\mu\nu}$ is the Minkowski metric. Similarly, when $\gamma_{\mu\nu}$
is a general Lorentzian metric and partial derivatives in (\ref{model}) are replaced by covariant ones, we obtain the isentropic Euler system on a general relativistic spacetime. Since the results in the present paper only concern the leading-order part of the basic field equation, these results apply to both special and general relativistic fluids. Let us however point out that our generalized model does not require any metric, so the only concept of derivative which is a priori available is that coming from the affine structure of $\mathbb{R}^4$, i.e. partial derivatives.\\
Provided that $\frac{\partial \rho}{\partial J^\alpha} J^\alpha$ is non-vanishing, the equations (\ref{model}) imply the conservation law
\begin{equation}\label{convext}
\partial_\mu J^\mu = 0.
\end{equation}
An outline of this paper is as follows. In Section 2 we begin with the concept of `convex extension' of a system of conservation laws
and its bearance on symmetric hyperbolicity. That is basically classical material due to \cite{fl}, but in a `covariant' setting, i.e.
without a given splitting of spacetime into space and time. In Section 3 we motivate the model given by (\ref{model}) in terms of standard special relativistic continuum mechanics and show that it generalizes the relativistic Euler system. Our main result concerning the existence of a convex extension and a symmetric hyperbolic form is stated in Section~4 and proven in Section~5. This result sheds new light on and generalizes the findings of
 \cite{frau,walton}. In Section~6 we sketch a uniqueness result related to weak solutions for which the conservation law $\partial_\mu J^\mu = 0$ is replaced by the inequality  $\partial_\mu J^\mu \leq 0$ across hypersurfaces of discontinuity.

\section{Symmetric hyperbolicity and convex extension}
This section is essentially a spacetime\footnote{We remark however that no metric is assumed here, so that no a priori concept of `timelike vector' (for instance) applies.} version of a classical result in \cite{fl}. We consider equations for variables $f = f^A (x^\mu)$ of the form
\begin{equation}\label{cons}
\partial_\mu\, U^\mu{}_A (f) = U^\mu{}_{A,B}(f) \,\partial_\mu f^B = 0,
\end{equation}
where $\mu = 0,1,...n$ with $n$ the number of spacetime dimensions and\footnote{For our specific model below, the role of the $f^A$-quantities is played by the field $J^\mu$ and the vector $V^\mu$ is simply $J^\mu$ itself.} $A=1,...N$. We also assume the convention that $W_{,A}$ is a partial derivative with respect to the variable appearing in parenthesis after the quantity $W$ ($f^A$ in the present case). All fields are assumed to be sufficiently smooth. We observe that it would be irrelevant for our purpose if zeroth-order terms in $f$ were present or $U^\mu{}_A$ would also depend on $x^\mu$. We assume that there exists a `non-characteristic covector', i.e.~a covector $\xi_\mu$ such that the vector $\eta_A (f) = U^\mu{}_A (f) \,\xi_\mu$ is an invertible function of $f$. If we introduce $\eta$ as a new dependent variable together with a coordinate system $(x^\mu)= (t,x^i)$ (with $i=1, \ldots, n$) such that $\xi_\mu dx^\mu = dt$
(where without loss of generality we assume that $\xi_\mu$ is chosen to be constant)
, then (\ref{cons}) takes the form
\begin{equation}\label{cons1}
\partial_t \eta_A + \partial_i \,\bar{U}^i{}_A(\eta) = 0,
\qquad
\bar{U}^i{}_A(\eta) = U^i{}_A(f(\eta)).
\end{equation}
This is the form in which systems of conservation laws are usually written (which we call the `standard form'), whereas we prefer to keep our equations in spacetime form. As will become clear below, the variable $\eta$, besides the condition that it exists, plays only a transitory role in our analysis.

Suppose next that (\ref{cons}) implies the equation
\begin{equation}\label{ext}
\partial_\mu V^\mu(f) = V^\mu{}_{,A}(f)\,\partial_\mu f^A = 0,
\end{equation}
in which the quantity $(V^0,V^i)$ is usually called an `entropy-entropy flux pair'. It follows from (\ref{cons}) that there exist quantities $g^A(f)$ such that
\begin{equation}\label{1}
g^C(f)\,U^\mu{}_{C,A} (f) = V^\mu{}_{,A}(f).
\end{equation}
To see this, we work in the standard form above and we write (\ref{ext}) as
\begin{equation}\label{ext1}
\partial_t \bar{V}^0 + \partial_i \bar{V}^i = \bar{V}^{0,A} \,\partial_t \eta_A + \bar{V}^{i,B} \,\partial_i \eta_B = 0.
\end{equation}
But for (\ref{ext1}) to hold identically for all functions satisfying (\ref{cons1}) it is necessary that
\begin{equation}\label{implies}
\bar{V}^{0,A}\bar{U}^i{}_A{}^{,B} = \bar{V}^{i,B}.
\end{equation}
Now setting $g^A(f) = \bar{V}^{0,A}(\eta(f))$, one checks that all components of (\ref{1}) are satisfied.

Eq.(\ref{1}) in turn implies
\begin{equation}\label{implies1}
U^\mu{}_{C,[A} (f) \,\,g^C{}_{,B]}(f)= 0
\end{equation}
and thus the system of equations
\begin{equation}
g^C{}_{,A}(f)\, \partial_\mu\, U^\mu{}_C (f) = g^C{}_{,A}(f)\, U^\mu{}_{C,B}(f) \,\partial_\mu f^B = 0
\end{equation}
is `symmetric'. If $g^A(f)$ is invertible, we can alternatively substitute $g$ for $f$ in (\ref{cons}) to obtain
the system of conservation laws
\begin{equation}\label{implies2}
\partial_\mu U^\mu{}_A (f(g)) = U^\mu{}_{A,C}(f) \,f^C{}_{,B}(g)\,\partial_\mu g^B = 0,
\end{equation}
which by simple linear algebra is again symmetric on grounds of Eq.(\ref{implies1}).

Eq.(\ref{1}) can be interpreted as saying that $g$ is the gradient of the quantity $\mathcal{U} := V^\mu \xi_\mu$ with respect to $\eta$. Furthermore we have
\begin{equation}
\mathcal{U}^{,AB} (\eta) = g^{A,B}(\eta) = (\eta_{A,B}(g))^{-1} = (U^\mu{}_{A,C}(f) \,f^C{}_{,B}(g) \xi_\mu)^{-1}.
\end{equation}
Thus definiteness of $\mathcal{U}^{,AB}$ is equivalent to that of $U^\mu{}_{C,A}\,f^C{}_{,B}\, \xi_\mu$. Now the symmetry of
$U^\mu{}_{A,C}(f) \,f^C{}_{,B}(g)$ together the with positive definiteness of $U^\mu{}_{A,C}(f) \,f^C{}_{,B}(g) \xi_\mu$ exactly says that
the second equation in (\ref{implies2}) is symmetric hyperbolic with subcharacteristic covector $\xi_\mu$. So
we have obtained the statement that the system of conservation laws (\ref{cons}), in terms of $g^A$, is symmetric hyperbolic with subcharacteristic covector $\xi_\mu$ if and only if $\mathcal{U}$ is a convex function of $\eta$.
The system just described is called a conservation law with convex extension $\mathcal{U}$. (Cf.~with the statements in \cite{da,l}.)

\section{Generalized relativistic hydrodynamics}
Many models of continuum mechanics, including elastic materials and perfect fluids in special relativity, can be written in the form
\begin{equation}\label{div}
\partial_\nu T_\mu{}^\nu = 0,
\end{equation}
where the tensor $T_\mu{}^\nu$ is constructed as follows: Let $\rho$, the `energy density', be a function of maps $F^A(x^\mu)$ from $\mathbb{R}^4$ into $\mathbb{R}^3$ of the form $\rho = \rho (F^A{}_\mu)$, where
$F^A{}_\mu = \partial_\mu F^A$, assumed to have maximal rank 3. Here $\mathbb{R}^4$, the collection of independent variables, plays the role of spacetime and $\mathbb{R}^3$ is the `material manifold', which should be thought of as an abstract collection of points labelling the particles making up the material continuum. Then the Euler-Lagrange equation for the action $S = \int \rho \,d^4x$ is equivalent to (\ref{div}) with
\begin{equation}
T_\mu{}^\nu = \rho \,\delta_\mu{}^\nu - \frac{\partial \rho}{\partial F^A{}_\nu}\, F^A{}_\mu.
\end{equation}
It is a system of four quasilinear second-order partial differential equations for the map $F$, of which three are independent,
since $\partial_\nu T_\mu{}^\nu = - F^A{}_\mu \partial_\nu \frac{\partial \rho}{\partial F^A{}_\nu}$.

Consider next the particle number current given by the spacetime vector
\begin{equation}\label{quant}
J^\mu = \Omega_{ABC}\,F^A{}_\nu F^B{}_\sigma F^C{}_\tau \,\epsilon^{\mu \nu \sigma \tau},
\end{equation}
where $\Omega$ and $\epsilon$ are respectively volume forms on $\mathbb{R}^3$ and $\mathbb{R}^{4 \star}$. The quantity
$J^\mu$ is conserved in the sense that
\begin{equation}\label{J}
\partial_\mu J^\mu = 0
\end{equation}
holds as a kinematical identity independently of the field equations.

We now consider a class of theories, which we call `generalized hydrodynamics' models, in which the Lagrangian $\rho$ is required to depend
on $F^A{}_\mu$ only via $J^\mu$.
Then, using the identity\footnote{For a consistency check, observe that (\ref{identity}) implies
$\frac{\partial J^\mu}{\partial F^A{}_{,\nu}} F^A{}_{,\nu}= 3 J^\mu$, which is nothing but the Euler identity for homogenous polynomials
of degree 3.}
\begin{equation}\label{identity}
\frac{\partial J^\mu}{\partial F^A{}_{\nu}} \,F^A{}_{\rho}= - 2 \,\delta_\rho{}^{[\mu} J^{\nu]},
\end{equation}
it follows that
\begin{equation}\label{T}
T_\mu{}^\nu = (\rho - \frac{\partial \rho}{\partial J^\alpha} J^\alpha)\, \delta_\mu{}^\nu + \frac{\partial \rho}
{\partial J^\mu} J^\nu,
\end{equation}
i.e. Eq.(\ref{new}).
We now change our viewpoint by regarding the equation $\partial_\nu T_\mu{}^\nu = 0$ as a system of four conservation laws
with the four components of $J^\mu$ being the dependent variables.

An important special case is where $\mathbb{R}^4$ is endowed with the Minkowski metric $\gamma_{\mu\nu}$ of signature
$(-+++)$ and $\rho$ depends on $J^\mu$, and $J^\mu$ only via
the quantity
\begin{equation}\label{n}
n = (- J^\mu J^\nu \gamma_{\mu\nu})^\frac{1}{2} > 0.
\end{equation}
In particular $J^\mu$ is assumed to be timelike. Then we have
\begin{equation}\label{mink}
T_\mu{}^\nu = \rho \,u_\mu u^\nu - p\,(\delta_\mu{}^\nu + u_\mu u^\nu),
\end{equation}
where $n u^\mu = J^\mu$, $u_\mu = \gamma_{\mu\nu} u^\nu$ and $p\,(n) = n \rho'(n) - \rho(n)$ with a prime denoting derivative with respect to $n$. In that case, the equation (\ref{div}) is the relativistic Euler equation for an isentropic perfect fluid. Similarly, when $\gamma$ is replaced by a curved Lorentz metric
$g$ and the derivative in (\ref{div}) is replaced by the covariant derivative with respect to $g$, we obtain the general relativistic Euler equations.

\section{Statement of the main result}

For the model proposed in Section 1, we can now state our main result.

\vskip.15cm

{\bf{Theorem.}}
{\sl  Define $B_\mu = \frac{\partial \rho}{\partial J^\mu}$ and $A_{\mu\nu} = \frac{\partial^2 \rho}{\partial J^\mu
\partial J^\nu}$, and let $(B,J) = B_\mu J^\mu$ be non-vanishing. Suppose furthermore that $A_{\mu\nu}$ has Lorentz\footnote{The
signature of the `generalized acoustic metric' $A_{\mu\nu}$ is in our convention minus that of spacetime.}
 signature $(+---)$
and
that $J^\mu$ is timelike with respect to $A_{\mu\nu}$, i.e. $A_{\mu\nu} J^\mu J^\nu = (J, AJ) > 0$. Then the system $\partial_\nu T_\mu{}^\nu = 0$, with
$T_\mu{}^\nu$ as in (\ref{T}), admits a convex extension given by $\mathcal{U} = J^\mu \xi_\mu$, where $\xi_\mu$ is any timelike covector
with respect to $A_{\mu\nu}$, that is, $(\xi, A^{- 1} \xi) > 0$ which is `future-pointing' in the sense that $(\xi,J)$ has the opposite sign as $(B,J)$.
}

\vskip.15cm

Importantly, in view of the discussion in Section~2, the symmetrization implied by this theorem guarantees the local well-posed\-ness of the initial value problem (with sufficiently regular initial data)
for the model of generalized fluids, which can thus be viewed, at least from a mathematical perspective, as a viable model for hydrodynamics.

Before proving this proposition we again look at the isentropic perfect-fluid case, where
\begin{equation}
(B,J) = n \rho'
\end{equation}
and
\begin{equation}
A_{\mu\nu} = - \frac{1}{n} \,\rho' \,(\gamma_{\mu\nu} + u_\mu u_\nu) + \rho'' u_\mu u_\nu.
\end{equation}
The quantity $A_{\mu\nu}$ is of course nothing but (a rescaling of) the acoustic metric of the fluid.
Our assumptions are equivalent to the conditions that $n \rho' = \rho + p$ and $\rho''$ both be positive. The first of these further implies that $n$, whence $p$ can be seen as a function of $\rho$. Consequently
\begin{equation}
\rho'' = \left(\frac{\rho + p}{n}\right)' = \frac{\rho + p}{n^2}\, \frac{d p}{d \rho} > 0.
\end{equation}
Thus we have that $\rho + p$ and $\frac{d p}{d \rho}$ have to be positive. The first condition is related to the dominant energy condition (DEC) for
perfect fluids - though the condition $\rho > 0$ is not needed. The second condition means that the sound velocity is real and positive. So our main Theorem, in the case of fluids, is valid provided that $\rho + p > 0$ and $\frac{d p}{d \rho} > 0$.

\section{Proof of the main result}

By a computation we find
\begin{equation}\label{TJ}
\hskip-1.9cm
\frac{\partial T_\mu{}^\nu}{\partial J^\sigma} = \frac{\partial \rho}{\partial J^\mu}\,\,\delta_\sigma{}^\nu
- 2 \left(\frac{\partial^2 \rho}{\partial J^\sigma \partial J^\tau}\right) J^{[\tau} \delta_\mu{}^{\nu]}
=B_\mu \,\delta_\sigma{}^\nu - 2 A_{\sigma \tau} J^{[\tau} \delta_\mu{}^{\nu]}, 
\end{equation}
which implies
\begin{equation}\label{TJ1}
J^\mu \frac{\partial T_\mu{}^\nu}{\partial J^\sigma} = (B,J) \,\delta_\sigma{}^\nu\,\,\,\Rightarrow\,\,J^\mu \,\partial_\nu T_\mu{}^\nu =
(B,J) \,\partial_\mu J^\mu.
\end{equation}
Thus, since $(B,J) \neq 0$, (\ref{div}) implies the conservation law
\begin{equation}\label{entails}
\partial_\mu J^\mu = 0,
\end{equation}
and the $V^\mu$ - quantity in (\ref{ext}) of section 2 is simply given by $V^\mu = J^\mu$.
The role of the $\eta_A$ - quantities of the previous section is now played by
\begin{equation}
\eta_\mu = T_\mu{}^\nu \xi_\nu.
\end{equation}
Provided that $(J,\xi)$ and $(\xi, A^{-1} \xi)$ are both non-vanishing, the Jacobian $\frac{\partial \eta_\mu}{\partial J^\nu}$
has an inverse. The proof is by computing $\frac{\partial \eta_\mu}{\partial J^\nu}$ and explicitly inverting the associated system of linear
equations. We omit a detailed proof, since we can directly read off the analogue of the $g^A$ - quantities of (\ref{1}), here called $w^\mu$, from the first equation in (\ref{TJ1}), written as
\begin{equation}
\frac{J^\mu}{(B,J)} \frac{\partial T_\mu{}^\nu}{\partial J^\sigma} = \frac{\partial V^\nu}{\partial J^\sigma}.
\end{equation}
Thus our system is symmetric in terms of the variable $w^\mu$ given by
\begin{equation}
w^\mu = \frac{J^\mu}{(B,J)},
\end{equation}
provided that the map sending $J^\mu$ to $w^\mu$ is invertible. To check this latter fact, we calculate
\begin{equation}
\frac{\partial w^\mu}{\partial J^\nu} = \frac{1}{(J,B)}\,\left[ \delta^\mu{}_\nu - \frac {J^\mu}{(J,B)}\, (B_\nu + A_{\nu \sigma} J^\sigma)\right],
\end{equation}
which has the inverse
\begin{equation}
\frac{\partial J^\sigma}{\partial w^\rho} = (J,B)\,\left[\delta^\sigma{}_\rho - \frac{J^\sigma(B_\rho + A_{\rho \alpha} J^\alpha)}{(AJ,J)}\right].
\end{equation}
Next we use $w$ as the new dependent variable in the partial differential equation given by $\partial_\nu T_\mu{}^\nu = 0$ and thus compute
$\frac{\partial T_\mu{}^\nu}{\partial J^\sigma}\frac{\partial J^\sigma}{\partial w^\rho} = \frac{\partial T_\mu{}^\nu}{\partial w^\rho}$
giving
$$ 
\frac{\partial T_\mu{}^\nu}{\partial w^\rho} 
= (J,B)\left[2 B_{(\mu} \delta_{\rho)}{}^\nu
+ \left(A_{\mu \rho} - \frac{2 B_{(\mu} A_{\rho) \sigma} J^\sigma + B_\mu B_\rho
+ A_{\mu \sigma} J^\sigma A_{\rho \tau} J^\tau}{(AJ,J)}\right) J^\nu\right].
$$ 
The symmetry of $\frac{\partial T_\mu{}^\nu}{\partial w^\rho}$ in $(\mu,\rho)$ is of course not accidental, see (\ref{implies2}) in the previous section.\\
Next, by our assumptions, the covector $\overline{\xi}_\mu = - (J,B)^{-1} A_{\mu\nu} J^\nu$ is future-pointing timelike with respect to $A_{\mu\nu}$. We now
compute
\begin{equation}\label{apply}
\frac{\partial T_\mu{}^\nu}{\partial w^\rho}\,\overline{\xi}_\nu = - (AJ,J)A_{\mu\rho} + B_\mu B_\rho + A_{\mu\sigma}J^\sigma
A_{\rho \tau}J^\tau,
\end{equation}
which in fact is positive definite. This is easily seen by applying the quadratic form given by the left-hand side of (\ref{apply}) to $J^\mu$
and any vector orthogonal to $J^\mu$ in the $A_{\mu\nu}$ - metric. Thus we have found one subcharacteristic covector, and so our system is symmetric hyperbolic
in the $w^\mu$ - variables. Since $\bar{\xi}$ is timelike with respect to $A$ and $(\bar{\xi},J)$ has the opposite sign of $(J,B)$, the statement of our proposition requires that we have to show positive definiteness of
$\frac{\partial T_\mu{}^\nu}{\partial w^\rho} \xi_\nu$ holds for any covector $\xi$ inside the same cone as $\overline{\xi}$.
For that purpose we first need to understand the structure of the characteristic variety $\mathcal{C}$. This is defined by the space of non-zero $\xi_\mu$'s for which
$\Delta(\xi) = \mathrm{det}  \left[\frac{\partial T_\mu{}^\nu}{\partial w^\rho}\, \xi_\nu\right]$ is zero. This can be computed with the result that
\begin{equation}
\Delta(\xi) = \frac{(J,B)^6
}{(AJ,J)}\,(J,\xi)^2\,(A^{-1} \xi,\xi) (- \mathrm{det} A).
\end{equation}
Thus the space of characteristic conormals is the union of two
(`future' and `past') conical hypersurfaces - the `sound cones' given by $(A^{-1} \xi,\xi) = 0$ - and a hyperplane outside these cones given by
 $(J,\xi) = 0$. The interior of the sound cones
are respectively two convex cones (in the sense that $\alpha \xi + \beta \eta$ for $\alpha \geq 0,\,\beta \geq 0$ belongs to these sets
when $\xi$ and $\eta$ do). Now let $\xi$ be future-pointing. It is then in the same convex cone as $\overline{\xi}$. This, in turn, is equivalent to
the polynomial in $\lambda$ given by $\Delta (\xi + \lambda \overline{\xi})$ having at most negative zeros. But zeros of this polynomial are
nothing but
the negative of the eigenvalues of the symmetric tensor $\frac{\partial T_\mu{}^\nu}{\partial w^\rho}\,\xi_\nu$, relative to the positive definite quadratic form given by $\frac{\partial T_\mu{}^\nu}{\partial w^\rho}\,\overline{\xi}_\nu$. Thus these eigenvalues (all of which have to be real) are positive, whence $\frac{\partial T_\mu{}^\nu}{\partial w^\rho}\,\xi_\nu$ is positive definite. This completes the proof of our theorem.

\section{Application}
In order for functions $f^A$ to (\ref{cons}) to have shocks, i.e.~be discontinuous across a hypersurface $\Sigma$, yet be a solution of (\ref{cons}) in the distribution sense, one finds by a standard argument that there has to hold
\begin{equation}\label{rh}
[U^\mu{}_A (f)] n_\mu = 0,
\end{equation}
where $n_\mu$ is the conormal of $\Sigma$, assumed to be everywhere non-characteristic, and $[\, \,]$ denotes the jump across $\Sigma$. The idea, then, in the theory of entropy solutions of (\ref{cons}) is to replace (\ref{ext})  by the condition that
\begin{equation}\label{ext2}
\partial_\mu V^\mu \leq 0
\end{equation}
hold in the distribution sense. In applications the conserved quantity associated with $V^\mu$ for strong solutions can e.g.~be negative entropy or energy.  In our case,
perhaps interestingly, it will be particle number.

Namely, consider the model of generalized hydrodynamics presented in Section~3 for which the condition (\ref{ext2}) reads
\begin{equation}\label{J-2}
\partial_\mu J^\mu \leq 0.
\end{equation}
By following the technique in \cite{da,dip} and taking advantage of the convexity property we established in our main theorem, we reach the following conclusion: if a smooth solution to the generalized hydrodynamics model exists, then it is unique even in the broad class of weak solutions, so long as it satisfies the inequality (\ref{J-2}).

\section*{Acknowledgments}

The first author (RB) was partially supported by Fonds zur F\"orderung der Wissenschaftlichen Forschung project P20414-N16. The second author (PLF) was partially supported by the Centre National de la Recherche Scientifique and the Agence Nationale de la Recherche through the grant ANR SIMI-1-003-01. RB and PLF also gratefully acknowledge financial support from the National Science Foundation under Grant No. 0932078 000 via the Mathematical Science Research Institute, Berkeley, where RB spent the month of October 2013 and PLF the Fall Semester 2013.

\end{document}